\documentclass[%
 reprint,
 amsmath,amssymb,
 aps,
onecolumn,pre]{revtex4}
\usepackage{amssymb}
\usepackage{amsmath}
\usepackage{graphicx}					
\usepackage{epstopdf}					
\usepackage[latin1]{inputenc}				

\usepackage{version}				
\usepackage{cases}						

\usepackage{mathrsfs}
\usepackage[squaren, Gray]{SIunits}			

\usepackage{color}
\usepackage{psfrag}
\usepackage{enumerate}
\usepackage{amssymb}
\usepackage{amsmath}
\usepackage{wasysym}
\usepackage{color}

\newbox\ncintdbox \newbox\ncinttbox
\setbox0=\hbox{$-$} \setbox2=\hbox{$\displaystyle\int$}
\setbox\ncintdbox=\hbox{\rlap{\hbox
    to \wd2{\kern-.1em\box2\relax\hfil}}\box0\kern.1em}
\setbox0=\hbox{$\vcenter{\hrule width 4pt}$}
\setbox2=\hbox{$\textstyle\int$}
\setbox\ncinttbox=\hbox{\rlap{\hbox
    to \wd2{\kern-.14em\box2\relax\hfil}}\box0\kern.1em}

\let\phi\varphi

\let\epsilon\varepsilon

\let\tilde\widetilde

\newcommand{\R}{{\mathbb R}}

\def\dd{\text{d}}

\def\B

\def\R{\ensuremath{\mathbb R}}

\def\P{\ensuremath{\mathcal P}}

\def\B{\ensuremath{\mathcal B}}

\def\L{\ensuremath{\mathcal L}}

\begin{document}

\title{Correlation dimension and phase space contraction via extreme value theory}

\author{Davide Faranda}
\email{davide.faranda@cea.fr}
\altaffiliation{Also at: London Mathematical Laboratory, 14 Buckingham Street, London, WC2N 6DF, UK}
\affiliation{%
LSCE-IPSL, CEA Saclay l'Orme des Merisiers, CNRS UMR 8212 CEA-CNRS-UVSQ, Universit\'e Paris-Saclay, 91191 Gif-sur-Yvette, France\\
}

\author{Sandro Vaienti}
\email{vaienti@cpt.univ-mrs.fr}
\affiliation{Aix Marseille Univ, Universit\'e de Toulon, CNRS, CPT, 13009, Marseille, France.}

\pacs{Valid PACS appear here}

\begin{abstract}
We show how to obtain theoretical and numerical estimates of correlation dimension and phase space contraction by using the extreme value theory. Maxima of  suitable observables sampled along the trajectory of a chaotic dynamical system converge asymptotically  to classical extreme value laws where: i) the inverse of the scale parameter gives the correlation dimension, ii) the extremal index is associated  to the   rate of phase space contraction for backward iteration, which in dimension $1$ and $2$ is closely related to the positive Lyapunov exponent and in higher dimensions is related to the metric entropy. We call it the Dynamical Extremal Index.  Numerical estimates are straightforward to obtain as they imply just a simple fit to an univariate distribution. Numerical tests range from low dimensional maps, to generalized Henon maps and climate data. The estimates of the indicators are particularly robust even with relatively short time series.
\end{abstract}

\maketitle

\section{Lead Paragraph}

\textbf{This study uses the link between extreme value laws and dynamical systems theory to show that important dynamical quantities as the correlation dimension, the entropy and the Lyapunov exponents can be obtained by fitting observables computed along a trajectory of a chaotic systems. All this information is contained in a newly defined Dynamical Extreme Index. Besides being mathematically well defined, it is almost numerically effortless to get as i) it does not require the specification of any additional parameter (e.g. embedding dimension, decorrelation time); ii) it does not suffer from the so-called curse of dimensionality. A numerical code for its computation is provided.
}
\section{introduction}

Since its introduction by Grassberger and Procaccia \cite{GB1, GB2}, the correlation dimension (CD) has been used as a powerful indicator for the description of  the fractal structure of invariant sets in dynamical systems. Similarly, the Lyapunov exponents and the entropy \cite{wolf1985determining,rosenstein1993practical} provide an indication of the relevant time scales associated to the dynamics and the predictability horizon of the system. Given the importance of these quantities, there exists an increasing body of literature on how to estimate CD, Lyapunov exponents and entropy.  It has been shown that reliable estimates of CD can be obtained with relatively short time series \cite{theiler1992testing}. Instead, the computations of Lyapunov exponents and entropy are still challenging because the existing methodologies require as input additional parameters as the dimension of the phase space and the relevant time scale of the dynamics (e.g. the decorrelation time). Calculations are then limited to the top Lyapunov exponent and  the reliability of estimates from time series of experimental phenomena is often questioned \cite{eckmann1992fundamental}. We defer the reader to the monographs  \cite{kantz2004nonlinear}, \cite{pikovsky2016lyapunov} and to the articles \cite{kantz2013problem}, \cite{politi2017quantifying} for recent advancements on the various statistical tools to investigate nonlinear time series.\\

The extreme value theory (EVT) has been used to characterize  the evolution of chaotic systems \cite{freitas2010hitting,faranda2011numerical}. It is possible to obtain dynamical properties in phase space (fractal dimension or stability) by exploiting the limiting theorems of the extreme value theory. The main idea is: i) to replace the stochastic processes used in the statistical framework with a trajectory of a chaotic dynamical system, ii) to study  the convergence of maxima of suitable observables to the classical extreme value laws. The parameters of the EVT provide estimates of dynamical properties of the system. This connection between EVT and the dynamical properties of chaotic systems is rich not only from a theoretical but also from a numerical perspective. Indeed the estimates of local properties obtained with EVT do not require the introduction of additional parameters and they are easy to implement numerically. They have been used to get insights on the dynamical behavior of atmospheric flows in \cite{faranda2013recurrence,faranda2016return,faranda2017dynamical}. In \cite{pons2017attractor}, it has been shown that the numerical algorithm based on EVT provide reliable estimates of the dimension of high dimensional systems up to phase spaces with thousands of dimension. It is  therefore desirable to estimate other key dynamical quantities in the EVT framework.\\

The purpose of this communication is to show that the correlation dimension  and the EVT are intimately related:  the CD arises by studying the distribution of the maxima of a new suitable observable evaluated along the orbit of a chaotic system. Moreover,  an exponent of the limit law, the extremal index,  is related, for hyperbolic attractors,   to the positive Lyapunov exponent  in dimension two and  to the metric entropy in  higher dimensions.  The idea of the relationship between EVT and CD comes from a previous work \cite{FGGV} where we used extreme value theory to detect and quantify the onset of synchronization in coupled map lattices. The relationship between the extremal index and the Lyapunov exponent and the entropy is new and  is particularly striking for maps with piecewise constant jacobian. In the general case, we derive a formula whose validity is confirmed by numerical experimets. We also explain the relation between our extremal index, the local dimensions and the phase space contraction. In the rest of the paper we will name it as the DEI, the {\em dynamical extremal index}.  We want to point out that our DEI is a  well defined quantity that can be used as a new indicator for the  sensitivity associated  to local hyperbolicity.   We will present the theoretical results in the next section: some of those results can be obtained by generalizing the techniques introduced in \cite{FGGV}; we will also address the need to develop a more appropriate theory of EVT for diffeomorphisms in higher dimensions.  We will then provide several examples of classical conceptual low-dimensional dynamical systems. We will discuss the implications of our results on higher dimensional  systems and the possibility to apply them to more general time series.  As an example, we will compute the indicators on climate data and explain how they provide a relevant physical information on the atmospheric circulation over the North Atlantic.

\section{Theoretical Results}
\subsection{A brief presentation of extreme value theory and a new observable}
Let $(M, \mu, T)$ be a dynamical systems given by a map $T$ acting on the metric compact  space $M$ with distance $\dd(\cdot, \cdot)$ and preserving the Borel  measure $\mu.$ Usually $M$ will be a compact subset of some $\mathbb{R}^n$ and $\dd$ a distance equivalent to the standard one. Let us take the {\em direct product}
$
(M\times M, \mu \times \mu, T\times T),
$
and denote with $(x,y)\in M \times M,$ a couple of point in the cartesian product $(M \times M).$ We then introduce the observable
$
\psi(x,y)=-\log \dd(x,y),
$
and consider the process $\{\psi\circ (T^j \times T^j)\}_{j\ge 0},$ and the maximum of the sequence
$
\mathcal{M}_n(x,y)=\max\{\psi(x,y), \psi(Tx, Ty),\cdots, \psi(T^{n-1}x, T^{n-1}y)\}
$
and finally its distribution $\mathbb{P}(\mathcal{M}_n\le u_n),$
where $\mathbb{P}=\mu \times \mu$ is the underlying probability and $u_n$ is a suitable scaling function tending to infinity and which we are going to define. Suppose that for a given positive number $\tau$ we can find a sequence of numbers $u_n$ such that
$
n \mathbb{P}(\psi\ge u_n)\rightarrow \tau, \ n\rightarrow \infty.
$
We say that the process $\{\psi\circ (T^j \times T^j)\}_{j\ge 0}$ satisfies an extreme value law of Gumbel's type if there is a number $\theta\in (0,1],$ the {\em extremal index}, such that
$
\mathbb{P}(\mathcal{M}_n\le u_n)\rightarrow e^{-\theta \tau}, \ \ n\rightarrow \infty.
$
We now introduce the diagonal neighborhood $S_n$ in the product space: $S_n=\{(x,y), \dd(x,y)\le e^{-u_n}\}.$  By substituting the expression of $\psi$ in $\mathbb{P}(\psi\ge u_n)$, we have
\begin{equation}\label{MU}
\mathbb{P}(\psi\ge u_n)=\mathbb{P}((x,y)\in S_n)=\int_M\mu(B(x, e^{-u_n}))d\mu(x),
\end{equation}
where $B(x, a)$ denotes the ball of radius $a$ centered on $x$ \footnote{Actually we got the equality of the right hand side in the limit of large $n$ when the two small corners of $S_n$ become negligible.}.
The quantity $\int_M\mu(B(x, r))d\mu(x)$ scales like $r^{D_2}$ and the exponent $D_2$ is called the {\em correlation dimension} and it characterizes the fractal structure of the support of $\mu$; a more formal, from the mathematical point of view, definition of this fact is given in \cite{P}, sect. 17, and references therein \footnote{A precise definition consists in taking the $\limsup$ and $\liminf$ of the ratio of the logarithm  with $\log(1/r).$}.  By injecting successively into (\ref{MU})  we have therefore that for large $n$:
\begin{equation}\label{FO}
u_n\sim \frac{-\log \tau }{D_2}+\frac{\log n}{D_2}:= \frac{z}{a_n}+b_n.
\end{equation}
where $\tau=e^{-z}$, $a_n=D_2$ and $b_n=\frac{\log n}{D_2}.$
For  numerical  purposes,   distribution functions like $\mathbb{P}(\mathcal{M}_n\le z)$ are modelled, for $n$ sufficiently large, by the so-called {\em generalized extreme value (GEV)} distribution which is a  function depending upon three parameters $\xi\in \mathbb{R}, \kappa\in \mathbb{R}, \sigma>0$ and such that:
$
F_{\text{GEV}}(z;\kappa,\sigma, \xi)=\exp\left\{-\left[1+\xi\left(\frac{z-\kappa}{\sigma}\right)\right]^{-1/\xi}\right\}.$

The parameter $\xi$ is called the tail index;   when its value is $0,$ the GEV corresponds to the Gumbel type. The parameter $\kappa$ is called the location parameter and $\sigma$ is the scale parameter: for $n$ large, the scaling constant $a_n$ is close to $\sigma^{-1}$ and $b_n$ is close to $\kappa.$ Therefore, if we could fit a limit law of Gumbel's type with suitable normalizing parameters $a_n$ and $b_n,$ we immediately get the correlation dimension. Such a technique was previously used  with a different observable, and it allowed to get the so-called {\em information dimension} $D_1(x),$  another fractal dimension which provides  the scaling of the measure of a ball around a given point $x,$ see \cite{book} and references therein. Although the information dimension depends on the point $x$, its value is the same for almost all the choices of $x$ with respect to the invariant measure and such an averaged valued, simply $D_1,$ is larger or equal to $D_2,$ see \cite{B} for an account on the different fractal dimensions. In particular if we denote with $d_H$ the Hausdorff dimension, we have $D_2\le D_1\le d_H.$
\subsection{The spectral approach with the new observable for conformal repellers}

Before showing our numerical simulations for the computation of the CD, let us argue how we get a Gumbel's type asymptotic distribution with an extremal index $\theta$ of dynamical meaning. First, we consider  one-dimensional dynamical systems generated by uniformly expanding maps with an invariant set which could be a Cantor set and equipped with mixing Gibbs measures. These systems are better known as {\em conformal repellers},  -- see  for instance \cite{PP} for a recent contribution -- whose  measures are characterized by a potential $\phi$ of type $\phi(x)=-\beta \log |T'(x)|$, where $T'$ denotes the derivative of $T$ and  $\beta\in \mathbb{R}.$ If we denote them as $\mu_{\beta},$ they are given by $h_{\beta}\nu_{\beta}$ where the density $h_{\beta}$ and the {\em conformal} measure $\nu_{\beta}$ are respectively the eigenfunctions of the transfer operator (Perron-Fr\"obenius)  and of its dual, both with eigenvalue $\lambda_{\beta}=e^{Q(\beta)},$ being $Q(\beta)$ the topological pressure. We remind that the transfer operator $\P_T$ for the map $T$ is defined, for an observable $f$ in some suitable Banach space $\mathcal{B}$ -- for instance the space of Lipschitz continuous functions -- by the duality relation: $\int \P_T f d\nu_{\beta}=\lambda_{\beta}\int f d\nu_{\beta}.$  We defer to the monograph \cite{GK} for an introduction to thermodynamic formalism.  The conformal measure verifies the property
$
\nu_{\beta}(TA)=\lambda_{\beta} \int_A e^{-\phi}d\nu_{\beta},
$
where $T$ is one-to-one over the measurable set $A$. A powerful method to investigate the distribution of our  process $\{\psi\circ (T^j \times T^j)\}_{j\ge 0}$ consists in perturbing the transfer operator $\P$ of the direct product $T\times T$. The key observation is that by repeatedly  using the duality relation  we can write  $\mathbb{P}(\mathcal{M}_n\le u_n)=\lambda_{\beta}^{-2n}\int \int \tilde{\P}^n_n(h_{\beta}(x)h_{\beta}(y))d\nu_{\beta}(x)d\nu_{\beta}(y),$ where the perturbed operator $\tilde{\P}_n$ is defined by acting on observables $f\in \mathcal{B}$, as $\tilde{\P}_n(f)= \P(f {\bf 1}_{S_n^c}),$ and $S_n=\{(x,y); d(x,y)\le e^{-u_n}\}.$
 When $n$ tends to infinity, the characteristic function of the complement of $S_n$, $\bf {1}_{S_n^c},$ goes to the identity and the operators $\P$ and $\tilde{\P}_n$ converge to each other in $\mathcal{B}.$ If the {\em unperturbed} operator $\P$ has a spectral gap, it allows exponential mixing for the observables in $\mathcal{B}$. This compensate the lack of independence of the process  $\{\psi\circ (T^j \times T^j)\}_{j\ge 0}.$ The same is true for the operator $\tilde{\P}_n$ and the maximal,  isolated, eigenvalue of $\P$, $\lambda_{\beta}^2,$ is close to that of $\tilde{\P}_n$, $\tilde{\lambda}_{\beta,n}^{(2)}.$ More precisely: $\tilde{\lambda}_{\beta,n}^{(2)}\sim \lambda_{\beta}^2-(1-\lambda_{\beta}^2 q_0)\mathbb{P}(S_n),$ where now  $\mathbb{P}=\mu_{\beta}\times \mu_{\beta}$. We will define the factor $q_0$ in a moment. The operator $\tilde{\P}_n$ now decomposes as the sum of a projection along the one dimensional eigenspace associated  to the eigenvalue $\tilde{\lambda}_{\beta,n}^{(2)}$ and an operator with a spectral radius exponentially decreasing to zero and which can be neglected in the limit of large $n$. This allows us to write  $\mathbb{P}(\mathcal{M}_n\le u_n)\sim \lambda_{\beta}^{-2n}\tilde{\lambda}_{\beta,n}^{(2)n}\int \int h_{\beta}(x)h_{\beta}(y))d\nu_{\beta,n}(x)d\nu_{\beta,n}(y),$ where $\nu_{\beta,n}$ is the conformal measure for the perturbed operator and the double integral on the right hand side converges to $1$ for $n\rightarrow \infty.$ Finally, we get by approximating $\tilde{\lambda}_{\beta,n}^{(2)}$ as above: $\mathbb{P}(\mathcal{M}_n\le u_n)\sim \left[1-\frac{(1-\lambda_{\beta}^2 q_0)\mathbb{P}(S_n)}{\lambda_{\beta}^{2}}\right]^n\sim \exp{\left[-\frac{(1-\lambda_{\beta}^2 q_0)\mathbb{P}(S_n)}{\lambda_{\beta}^{2}}n\right]}.$ We now remind that we are under the assumption that $n\mathbb{P}(\psi\ge u_n)=n \mathbb{P}(S_n)\rightarrow \tau, n\rightarrow \infty.$ This lead to the Gumbel law $e^{-\theta \tau}$ provided  that the dynamical extremal index $\theta$ is defined as
\begin{equation}\label{EI}
\theta= \frac{1-\lambda_{\beta}^{-2} q_0}{\lambda_{\beta}^2}.
\end{equation}
The term $q_0$ is obtained by the previous perturbation theory under the assumption that the diagonal in the product space is left invariant by the direct product of the two maps. In particular we have:
\begin{equation}\label{Q0}
q_0=  \lim_{n\rightarrow \infty}\frac{\mathbb{P}(S_n\cap \overline{T}^{-1} S_n)}{\mathbb{P}(S_n)},
\end{equation}
provided that the limit exists. The technique just described was firstly proposed by Keller \cite{KE} as an alternative way to get EVT for systems with exponential mixing and it is based on a perturbative result by Keller and Liverani \cite{keller2009rare}.  We defer to \cite{KE} and to our paper \cite{FGGV} for a detailed presentation of that theory. It can be applied to conformal mixing repellers and it provides  the preceding estimates, namely the asymptotic scaling for the maximal eigenvalue.  We would like to point out that with our choice for the observable $\psi,$ the perturbative approach just sketched gives the Gumbel's law in a very direct and natural manner.

The computation of $q_0$ proceeds now as in \cite{FGGV} with a substantial difference: the nature of the conformal measure  does not imply necessarily that the  ratio $\frac{\nu_{\beta}(B(Tx, r))}{\nu_{\beta}(B(x, r))}$ is constant, which happened when the conformal measure was Lebesgue. This difficulty could be partially overcame by supposing that the potential  is constant, otherwise we could bound $q_0$ from above and below with (close) approximations of the potential. By assuming  that the latter is constant and equal to $\overline{\phi}$ and also that the density $h_{\beta}$ does not vary too much, we get that $q_0$ is of order $e^{\overline{\phi}}$ and therefore
\begin{equation}
\theta\sim  \frac{1-\lambda_{\beta}^{-2} e^{\overline{\phi}}
}{\lambda_{\beta}^2}.
\end{equation}
It is worth mentioning that whenever the conformal measure is Lebesgue ($\beta=1$), the above computation {\em can be made rigorous} as in Proposition (5.3) in \cite{FGGV} and it gives
\begin{equation}\label{FO}
\theta= 1-\frac{\int_M \frac{h^2(x)}{|T'(x)|}dx }{\int_M h^2(x) dx},
\end{equation}
where $h$ is the density of the invariant measure: we defer to our paper \cite{FGGV} for the assumptions on the system  which permit  to get such a result. In particular those systems contains conformal repellers with finitely many branches and an absolutely continuous conformal measures. Notice that by introducing the invariant measure $\mu=h dm$ we could identically write
\begin{equation}\label{je}
\theta= 1-\frac{\int_M h(x)e^{-\log|T'(x)|}d\mu(x)}{\int_M h(x) d\mu(x)}.
\end{equation}
If the derivative does not change too much we get $\theta\sim 1-e^{-\Lambda_{\mu}},$ where $\Lambda_{\mu}$ is the positive Lyapunov exponent of the measure $\mu.$ Alternatively, if the density $h$ could be considered constant we can bound (\ref{je}) by Jensen's inequality as
$$
\theta\sim 1-\int_M\frac{1}{|T'(x)|}d\mu(x)\le  1-e^{-\int_M \log|T'(x)|d\mu(x)}=1-e^{-\Lambda_{\mu}}.
$$
In both cases the DEI $\theta$ is related to the positive Lyapunov exponent: this analogy will be pursued in the next section.
\subsection{Attractors and high dimensional systems}

For invertible maps generating attractors endowed with the SRB measure, the computation of   the dynamical extremal index  is less straightforward; we should stress that a spectral theory of extreme value for (invertible) uniformly hyperbolic maps is still missing.  Suppose we  take an hyperbolic diffeomorphisms $T$ preserving the ergodic  SRB measure $\L$. Then the quantity $q_0$ in (\ref{Q0}) becomes
\begin{equation}
q_0=\lim_{n\rightarrow \infty}\frac{\int d\L(x) \int {\bf 1}_{S_n}(x, T^{-1}y){\bf 1}_{S_n}(Tx, y)d\L(y)}{\int d\L(x) \int {\bf 1}_{S_n}(x,y)d\L(y)}.
\end{equation}
When we iterate backward the points $y\in B(Tx, e^{-u_n}),$ we should keep only those points whose preimage is at a distance at most $e^{-u_n}$ from $x.$ Those preimages form a set $Q(x)$ which is obtained by squeezing the ball $B(Tx, e^{-u_n})$ along the unstable manifolds.  Let us suppose that the tangent expanding subspace $\Sigma_u(Tx)$ at $x$   has dimension $d.$  Then the  measure of  $Q(x),$ and therefore, by the forward invariance of the measure, of its image in $B(Tx, e^{-u_n})$ will be of order $|\det (DT(x)|_{u})|^{-1} \ \L (B(Tx, e^{-u_n}))$, where $DT(x)|_{u}$ is the  derivative of $T$ restricted to $\Sigma_u(x)$. We remember in fact that the conditional SRB measure   on the unstable manifolds is smooth. This immediately gives $q_0$ of order
\begin{equation}\label{pe}
q_0\sim \frac{ \int d\L(x) |\det (DT(x)|_{u})|^{-1}\ \L (B(T(x), e^{-u_n})) }{\int\  d\L(x) \ \L (B(x, e^{-u_n}))}.
\end{equation}
We see that $q_0$ contains information about the dimension through the scaling of the denominator; we are now interested in  the contribution of the other term in the numerator. At this regard we first  remind that, for SRB measures, we can use the Pesin's formula \cite{Viana}:
$$
\int d\L(x) |\det (DT(x)|_{u})|=\sum_{j=1}^d \Lambda^+_j=h_{\L},
$$
where the $\Lambda_j^+$ are the positive Lyapunov exponents  with multiplicity one, and $h_{\L}$ is the metric entropy of the SRB measure. We now proceed under two assumptions as we did at the end of the previous section.  Let us first  assume that the derivative along the unstable subspaces does not vary too much. Then,  we could estimate the DEI as
\begin{equation}\label{GE0}
\theta\sim 1-e^{-h_{\L}}.
\end{equation}
For $d=1$ we can replace the entropy with the (unique) positive Lyapunov exponent $\Lambda_{\L};$ in the following we will simply write it as $\Lambda_+.$\\ The other assumption exploits the fact that for these system and for $\L$-almost all points $x$ we have, by Young's theorem \cite{LSY}, that $\lim_{r\rightarrow 0}\frac{\log\L(B(x,r))}{\log r}=D_1,$ where $D_1$ is the information dimension. Hence we could guess that $\L (B(x, e^{-u_n}))\sim e^{-{u_n}D_1}$ and therefore forget about the dependence on the variable $x$. This is generally  false since the multiplicative factor in the previous scaling could depend on $x$. Indeed  when we integrate  $\L (B(x, e^{-u_n}))$ we get $D_2$ which could be different from $D_1$. If we suppose that the dependence on $x$ of the prefactors is negligible, which means that we are considering an homogenous fractal invariant set with $D_1\sim D_2,$ then we have for the DEI:
\begin{equation}\label{GE0}
\theta  \sim 1-\int d\L |\det(DT(x)|_u)|^{-1}\le   1-e^{-\int d\L(x) |\det (DT(x)|_{u})|} = 1-e^{-h_{\L}},
\end{equation}
where the derivative is {\em not}  supposed to be constant and where we have used again the Jensen's inequality to establish the upper bound.

 Those two  approximations are very crude; we are in fact either neglecting the contributions of the prefactors in the local scaling of the balls in (\ref{pe}), or not taking into account   the geometric factors when  the ball $B(Tx, e^{-u_n})$ is squeezed at a distance $e^{-u_n}$ from $x.$ Moreover, the variation of the derivative, especially sensible in the non-uniformly hyperbolic setting,  could give large differences in the determination of the DEI, as we experience for instance for the H\'enon map, see below. The preceding relation
is pretty well satisfied for maps with with one-dimensional unstable subspace and (piecewise) constant jacobian, like  the Baker transformation, the Lozi map and the solenoid.  For  the algebraic automorphism of the torus (cat's map), a simple  argument allows us to improve the previous rate just by taking into account the geometric factors.  Surprisingly, relation (\ref{GE0}) is pretty well satisfied  in the example below of the generalized H\'enon maps, where  the unstable subspace has dimension larger than one, i.e. we have more than one  positive Lyapunov exponent. In conclusion our index $\theta$ traces in a satisfactory way the entropy. The eventual deviations  are due to the variation of the derivative and the local scaling of balls in (\ref{pe}). Although These effects are difficult to compute analytically, the DEI $\theta$ is relatively easy to compute numerically and it furnishes a new indicator for the local instability in chaotic systems.

\section{Numerical Computations}
The numerical computations presented in the remaining of this work are performed by using the numerical algorithms and codes detailed in the supplemental material. The stability of the results is checked against different $l,n,m,s$. In particular, we perform two sets of simulations. The first set of accurate simulations consist of $l=100$ trajectories, with $n=10^6$ iterations, $m=10^3$ blocks of $s=10^3$ length each. The second set of $l=100$ simulations consists of short series of $n=10^4$ iterations, with $s=m=10^2$. This second set is useful to check whether the technique is reliable also for short time series.  Except where specified, we use  $\tilde{s}=0.99$ for the following computations. However, results are stable when considering different quantiles ranging from $0.97<\tilde{s}<0.999$.

\subsection{Low dimensional maps}

We begin the numerical computations with several examples on low dimensional maps. A summary of the results for all maps analysed is reported in Table 1. For a few maps, we report the model equations in the supplemental material to streamline the exposition.

\begin{itemize}
\item Let us begin with  the Bernoulli Shift map $T(x)=3 x$-mod $1.$ For this system $D_2=1$ and $\theta=1-1/3=2/3$. The numerical estimates (Table 1) are coherent with the theoretical values for both accurate and short simulations.

\item We now consider the Gauss map $T(x)=\frac{1}{x}$-mod$1$ defined on the unit interval. Although, strictly speaking, this map does not fit the assumptions in \cite{FGGV} since in the latter paper we consider maps with finitely many branches, we still try formula (\ref{FO}). For the Gauss map the density is explicit and reads $h(x)=\frac{1}{\log2}\frac{1}{1+x}.$ The integral in (\ref{FO}) can be easily computed and gives $\theta= 4\log(2)-2 \sim 0.77$, whereas $D_2$ is expected to be 1. The numerical estimates are coherent with the theoretical values (Table 1).

\item Returning to a map with constant slope $3$, we now look at the transformation generating the classical ternary Cantor set.  In order to compute numerically the GEV function, one should access  the invariant Cantor set, which is of zero Lebesgue measure. We need therefore to  use the backward iterates of the map (otherwise almost all the forward orbits will fall into the holes), and the measures allowing us to compute the time averages are the so-called {\em balanced measures}, given suitable weights to the preimages of the map: see our article \cite{LET}, section 3.2.2 for a description of such measures. For the ternary Cantor set and choosing equal weights $1/2$ for the two preimages, it is easy to check that such a balanced measure coincides with the Gibbs measure with $\beta=\log2/\log3$ which is  the Hausdorff dimension  of  the invariant set. The measure $\mu_{d_H}$ is called {\em uniform}, see \cite{B}, Sect.3. The potential $\phi$ will be equal to $-\log 2$ and $\lambda=1$, since by Bowen's formula $Q(d_H)=0.$ Therefore for the ternary Cantor set we get a DEI equal to $0.5$ which is perfectly confirmed by the numerical simulations (Table 1).

\item For the Lozi  map : $x_{n+1}=a|x_n|+y_n+1, y_{n+1}=bx_n, \ a=1.7, \ b=0.5,$
 $\Lambda_+$ is of order  $0,47$ \cite{PV},  which gives, with our approximation, a DEI of order $\theta=0.37$. Previous numerical computations for $D_2$ gave $D_2\sim 1.38,$ \cite{SR}. Our computations (Table 1) are coherent with the theoretical values.
\item For the H\'enon  map   $x_{n+1}=ax_n^2+y_n+1, y_{n+1}=bx_n, \ a=1.4, \ b=0.3,$ $\Lambda_+$ is of order $0,42$ \cite{PV}, which gives, with our approximation, a DEI of order $\theta=0.34$. Previous numerical computations for $D_2$ gave $D_2\sim 1.22,$ \cite{SR}. The GEV computations give  $D_2=1.24\pm0.11$ but $\theta=$0.43$\pm$0.01 for $n=10^6$ (See Table 1 for the results with $n=10^4$ iterations). The discrepancy of the DEI estimate does not get any better with the increase of $\tilde{s}$ or $n$. As said before, we do not expect $\theta$ to coincide with the estimate $0.34$  due to the variation of the derivative and the non-uniform hyperbolicity of the map.
\item Let us consider the cat's map  with the associated matrix $
\quad
\begin{pmatrix}
1 & 1 \\
1 & 2
\end{pmatrix}.
$
The stable and unstable manifolds for such a map are orthogonal, so we could suppose that the preimage of the ball $B(Tx, e^{-u_n})$ will intersect the ball $B(x, e^{-u_n}) $   in a rectangle $R(x)$ centered at $x$ and with the shortest side of length $(\lambda_+)^{-1} e^{-u_n},$ where $\lambda_+= \frac{3+\sqrt{5}}{2}$ is the eigenvalue larger than $1$ corresponding to the unstable direction. An elementary calculation shows immediately that $q_0\sim \L(R(x))/\L(B(x, e^{-u_n})$ is approximately given by $\frac{4}{\pi}(\lambda_+)^{-1}$ which gives an extremal index as $0.51.$  Previous numerical computations for $D_2$ gave $D_2\sim 1.987,$ \cite{SR}. The numerical computation with the GEV fitting gives
$D_2=2.00\pm0.06$ and $\theta=0.552\pm0.005$ for $n=10^6$.
 In order to investigate the discrepancy with our theoretical estimate, we raised the quantile from $\tilde{s}=0.99$ to $\tilde{s}=0.999$, i.e. we select more extreme clusters. The estimates for this case are $\theta=0.54\pm0.02$, more compatible with the theoretical one. Finally, if we consider longer trajectories ($n=10^7$ iterates) with an even higher quantile ($\tilde{s}=0.9999$), we get $\theta=0.53\pm0.06$, which is even closer to   the theoretical guess.
\item We now consider the baker's map (see supplemental material); it depends on three parameters $\alpha, \gamma_a, \gamma_b.$
The positive Lyapunov exponent is given by \cite{B}, eq. 5.14:
$$
\Lambda_+=\alpha\log\frac{1}{\alpha}+(1-\alpha)\log\frac{1}{1-\alpha}
$$
With the value $\alpha=1/3, \gamma_a=1/5, \gamma_b=1/4,$ we get $\Lambda_+\sim 0,64$ which gives, with our approximation,  an extremal index of order $0,47$.  In the paper \cite{B}, eq. 5.18, we gave an implicit formula expressing $D_2$ as a function of $\alpha$ and with respect to the SRB measure. For $\alpha=1/3$, this estimate reads $D_2\simeq 1.41$. The GEV estimates are  given in Table 1 and are consistent with the theory. \\

    \item We next consider an attractor embedded in $\mathbb{R}^3,$ the so-called solenoid, see supplemental material; it depends upon the parameter $a\in (0, 0.5).$   The attractor is foliated by one-dimensional unstable manifolds, while each meridional disk  is a two-dimensional stable manifold each of which intersecting the attractor over a Cantor set. The Lyapunov exponents are
        $$
        \Lambda_{-}=\log a<0, \ \Lambda_{+}=\log 2,
        $$
        while the Hausdorff dimension $d_H$ is given by the formula \cite{simon1997hausdorff}
        $$
        d_H=1+\frac{\log2}{-\log a}.
        $$

The numerical computations for the solenoid provide a further test of the validity of the numerical algorithm and are provided in Table 1.
 \end{itemize}

\subsection{High dimensional generalized H\'enon maps}

\begin{figure}
\begin{center}
\includegraphics[width=.69\textwidth]{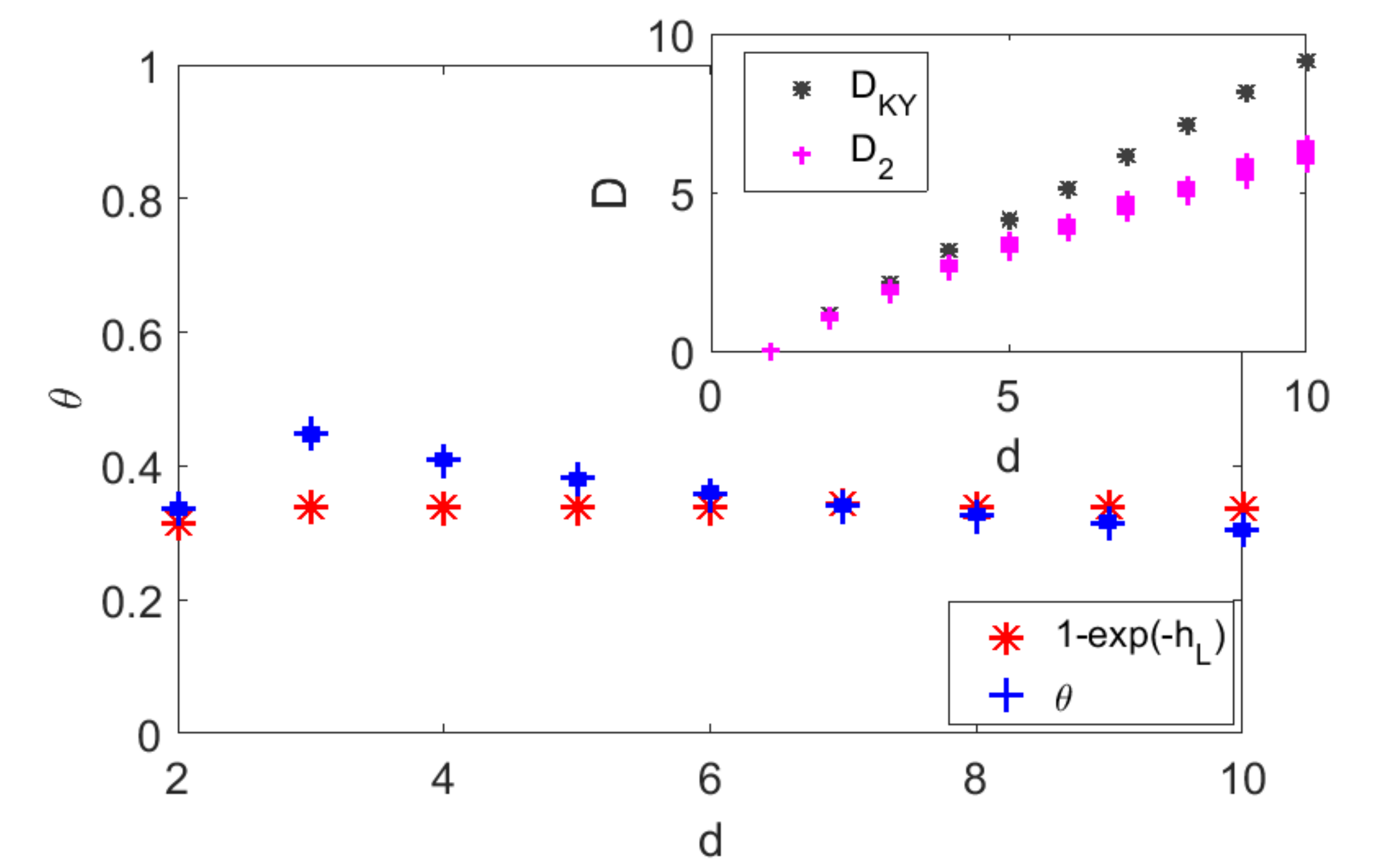}
\caption{Estimates of the dynamical extremal index $\theta$  and correlation dimension $D_2$ (inset) obtained for the Generalized Henon maps (Eq. \ref{HHM} in different dimensions $d$. Values represent  the estimates obtained taking 30 couples of trajectories, iterated for $n=10^6$ iterations. Each couple is displayed using a single marker, but the uncertainty is so small that difference between couples are hardly recognizable. The quantile used for the estimation is $\tilde{s}=0.98$. Results are compared to those obtained using the Kaplan-Yorke dimension $D_{KY}$ and the entropy $h_{\L}$. This map has $d-1$ positive Lyapunov exponents.}
\label{henon}
\end{center}
\end{figure}

We now analyze the generalized H\'enon maps defined in \citep{baier1990maximum} and further analyzed in \citep{richter2002generalized}. They are defined as:

\begin{equation}
x_{n+1}(1)=ax_n(d-1)^2-bx_n(d) \qquad x_{n+1}(i)=x_n(i-1)
\label{HHM}
\end{equation}
 When the parameter $a=1.76$ the number of positive Lyapunov exponents is $d-1$; we could therefore test our relation (\ref{GE0}) by computing the entropy $h_{\L}$ as the sum of positive  Lyapunov exponents (see Table 2 in \cite{richter2002generalized}) for a given $d$. We also perform the computation of the dimension $D_2$ and compare it to the Kaplan-Yorke dimension $D_{KY}$ given in \cite{richter2002generalized}; we used such a dimension because we did not find an explicit computation of $D_2$ in the literature.  The good agreement between our numerical results (Figure \ref{henon}) confirm the validity of Eq. \ref{GE0} with the caveat that an exact correspondence cannot be derived for the geometric factor that stretch balls in phase space in different dimensions: the origin of this discrepancy has been discussed in detail at the end of Section 3.3.

\subsection{Application to atmospheric data}

We now consider an application to atmospheric data. The purpose of this application is to show that the applicability of the technique on real data  provides results that have a coherent interpretation in terms of the underlying physics of the systems. In order to provide evidence of the robustness of our results, we will study several trajectories of a climate models which incorporate observations of the past 110 years, and compute $\theta$ and $D_2$ for several sub-periods showing that the results are numerically stable. We study the atmospheric circulation over the North Atlantic and focus on a single field that represents its major features: the sea-level pressure (SLP) \citep{Hurrell676, moore2013multidecadal}.   Indeed, it has been shown that  SLP fields can  be used to study teleconnection patterns as well as storm track activity and atmospheric blocking \citep{rogers1997,comas2014impacts}. The trajectories of our dynamical systems are  successions of  SLP fields extracted with daily frequency from  the ERA-20CM reanalysis project  over the period 1900-2010 \cite{hersbach2015era}. The ERA-20CM consists of 10 members ensemble of  a (climate) model whose task is to reconstruct at best the 1900-2010 atmospheric dynamics by constraining the model to include the information  from available surface observations. Each member of the ERA 20CM is therefore a slightly perturbed reconstruction of the atmospheric dynamics in the past 110 years.  The choice of the North Atlantic domain (80$^\circ$ W $\leq$ Long. $\leq$ 50$^\circ$ E, 22.5$^\circ$ N$ \leq$ Lat. $\leq$  70$^\circ$ N ) is motivated by the better observational coverage over the region in the first part of the analysis period compared to other regions of the globe \citep{krueger2013inconsistencies}. Before presenting the results for $D_2$ and $\theta$ we would like to stress that i) our analysis will  only be representative of the North-Atlantic domain and $D_2$ will be a proxy of the active degrees of freedom of the atmospheric circulation in this area. Therefore, our results  cannot be used to estimate the dimension of the full atmospheric climate attractor. ii) Previous results \cite{faranda2017dynamical, rodrigues2017dynamical,faranda2017npg} have shown that the estimates obtained for the daily dimensions are robust with respect to the changes in the datasets, resolution of the climate models, and are linearly insensitive to the size of the domain. This gives us confidence on the applicability of the numerical algorithm described in this paper for climate data since it is largely based on those used in \cite{faranda2017dynamical, rodrigues2017dynamical,faranda2017npg}.\\

\begin{figure}
\begin{center}
\includegraphics[width=.69\textwidth]{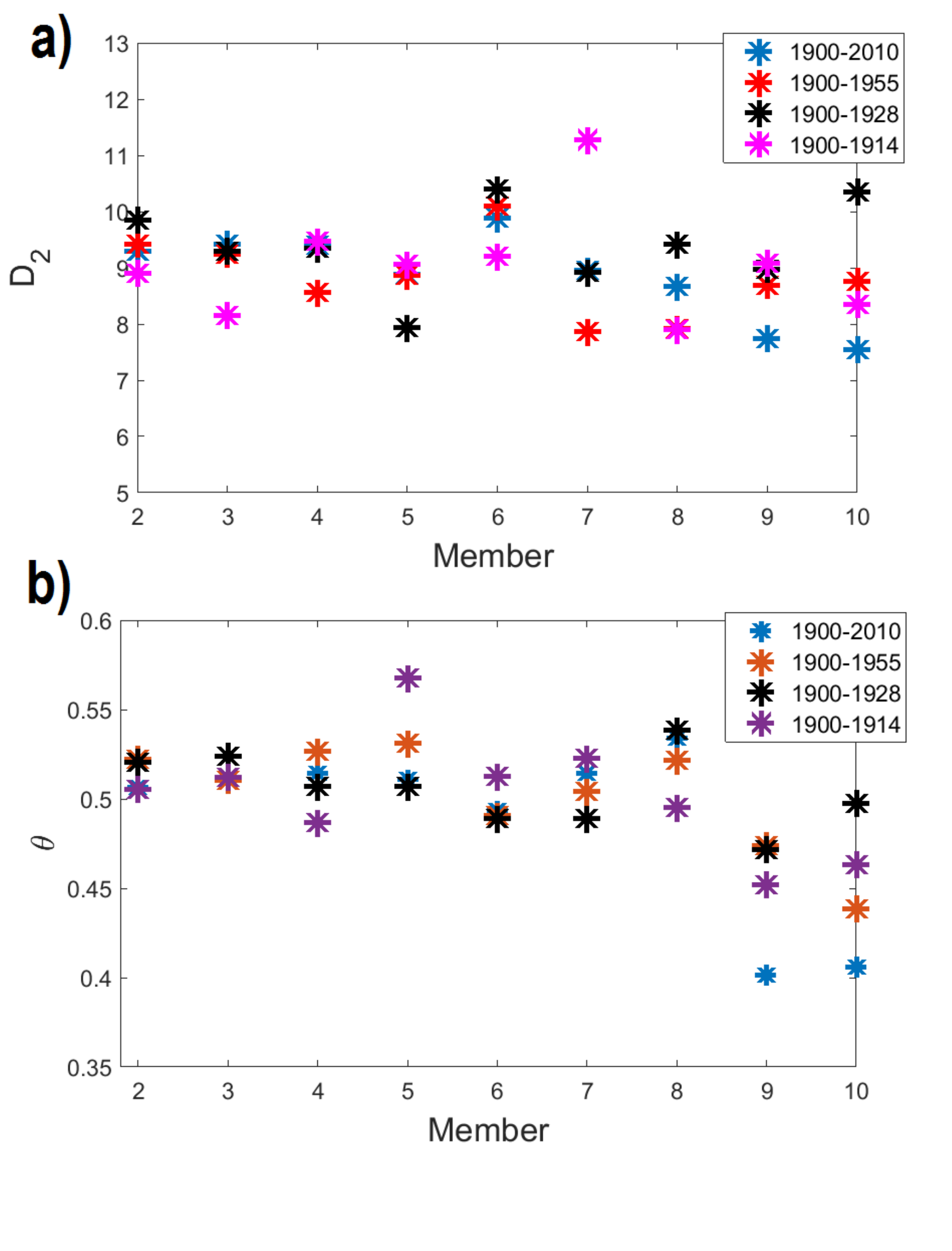}
\caption{Estimates of correlation dimension $D_2$ (a) and extremal index $\theta$ (b) obtained for daily sea-level pressure maps for four different periods in the  ERA-20CM reanalysis. Values represent   the estimates obtained taking as reference trajectory $x$ the member M1 and as $y$, the remaining 9 ensemble members.}
\label{climate}
\end{center}
\end{figure}

The result for $D_2$ and $\theta$ on the SLP fields of the ERA-20CM ensemble are presented in Figure \ref{climate}. For each estimate, we fix the reference trajectory $x$ as the first member (M1) of the ERA-20CM ensemble because this is always considered as the reference simulation, while $y$ is alternatevely set as the M$i$th member with $i=2,3,\dots,10$.  The dependence of the results on the reference member are tested in the supplemental material figure S1. To test the robustness of the results, we provide four  estimates of $D_2$ and $\theta$: i)  using the full data in the period 1900-2010, ii) using 1900-1955 data, iii) using 1900-1928 data and iv) considering only the first 14 years (1900-1914) of data. For each member, the results are reported in Figure \ref{climate}. The ensemble averages of $D_2$ and $\theta$  for the different periods are instead reported in Table 2. Estimates are consistent for different periods and the value of $D_2\simeq9$ found on average, is slightly lower than the estimates of $d_H$ found in \cite{faranda2017dynamical} (we remind that $D_2<d_H$). The value of $D_2$ roughly corresponds to the number of spatial degrees of freedom active in a North-Atlantic SLP field as explained in \cite{faranda2017dynamical}. Indeed, the domain used for this analysis can host about 9 large spatial stuctures reparted between 3-4 extratropical cyclones at time and the same number of anticyclones (see the textbook of Holton \cite{holton} for estimates of the typical size of these objects). $\theta$ is, in fact, the inverse of the average time the two trajectories $x$ and $y$ cluster togheter. The value of the DEI $\theta=0.5$ corresponds therefore to a contraction of the phase space associated with a timescale between 2 and 3 days.  This is the typical decay rate of baroclinic eddies associated to the low pressure systems observed in SLP fields (see again the textbook by Holton \cite{holton} for the decay rates). We finally notice that our formula (\ref{GE0}) gives for the entropy the value $\log 2.$  In Figure S2, we show a moving window computation of $D_2$ and $\theta$. No clear trend emerges that could be attributed to anthropogenic focing. This result is consistent with those found for $d_H$ in \cite{rodrigues2017dynamical}. We remark however some differences in the variability of the indicators among the members. In particular, M9 and M10 have a minimum of $\theta$ around 1960. This could be due to the different boundary conditions applied to the members and detailed in \cite{hersbach2015era}.

\subsection{Additive noise}

\begin{figure}
\begin{center}
\includegraphics[width=.69\textwidth]{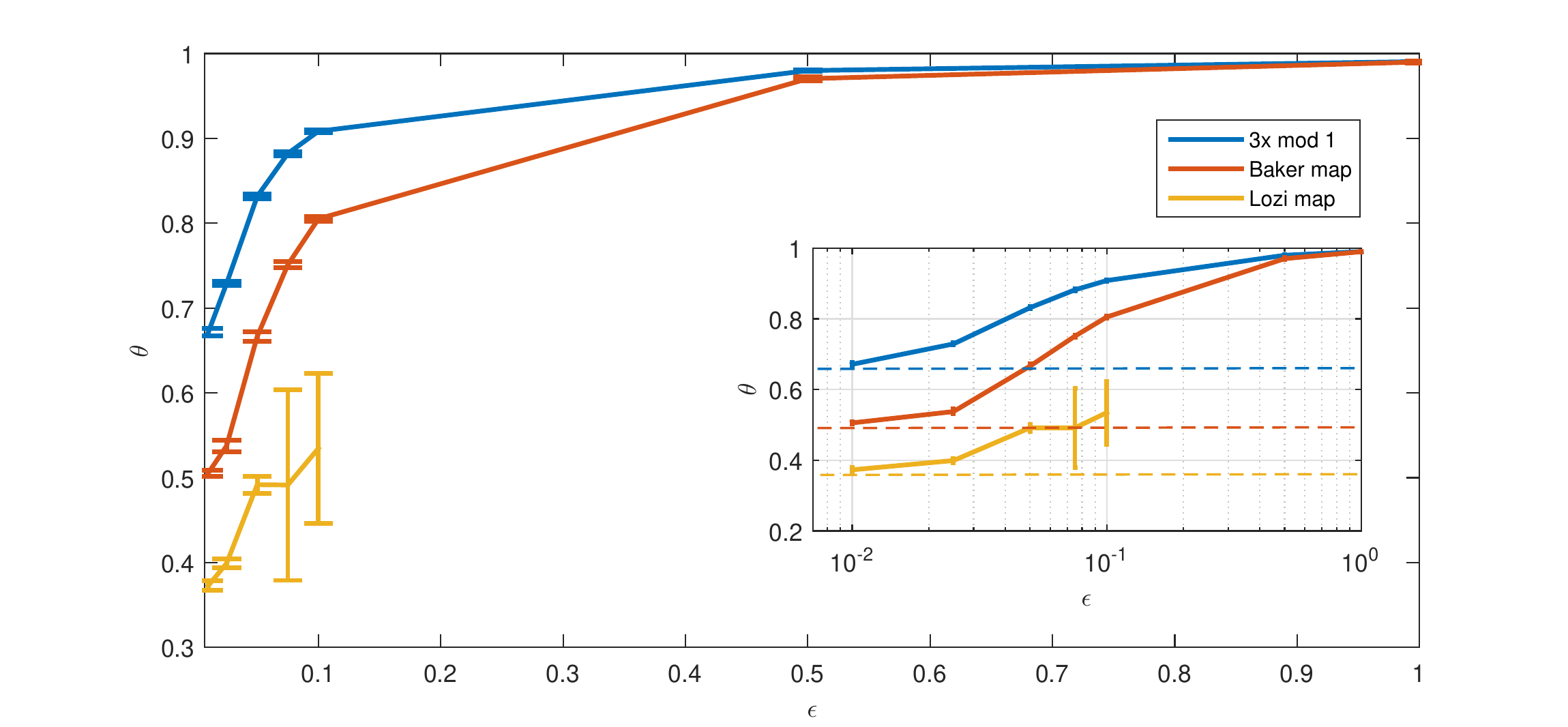}
\caption{Dynamical extremal index $\theta$ vs intensity of the additive noise $\epsilon$ for three different maps: 3x mod 1 (blue), Baker map (red) and Lozi map (orange). The errorbar indicates the standard deviation of the sample of $l=100$ trajectories, each consisting of $n=10^6$ iterations.  The quantile for the estimate of the extremal index is $\tilde{s}=0.99$. The inset shows the same data in semilog scale, with the deterministic values represented by dotted lines.}
\label{ExtI}
\end{center}
\end{figure}

In our previous papers \cite{faranda2013extreme}, \cite{aytacc2015laws},\cite{FGGV} we have analyzed the effect of additive noise on the parameters of the extreme value laws. It consists in
defining a family of maps $T_{\xi}=T+\epsilon \xi$ with $\xi$ a random variable sampled from some  distribution $\mathbb{G}$ (we will take here the uniform distribution on some small ball of radius $\epsilon$ around $0$). The iteration of the single  map $T$  will be now replaced by the concatenation $T_{\xi_n}\circ T_{\xi_{n-1}}\cdots\circ T_{\xi_1}$ and the evaluation of an observable computed along this orbit will be given by the probability measure $\mathbb{P}$ which is the product of $\mathbb{G}^{\mathbb{N}}$ with
 the so-called {\em stationary measure} $\mu_{S}$, verifying,  for any real measurable bounded function $f$: $\int f d\mu_s=\int f\circ T_{\xi}d\mu_s:$  see \cite{book} Chapt. 7 for a general introduction to the matter.
In the  aforementioned papers  \cite{faranda2013extreme} and  \cite{aytacc2015laws}
we have shown analytically that for  dynamical systems perturbed additively the extremal index $\theta=1,$ no matter what the intensity of the noise is. The proof was supported by numerical experiments, using also  different noise types. The
extremal index  is a parameter that quantifies the
amount of clustering, the stickiness of the trajectory in phase space. In our setting,  clustering happens in presence of invariant sets, which are periodic points in \cite{faranda2013extreme}. By looking at formula (\ref{Q0}), we see that we estimate the proportion of the neighborhood of the invariant set returning to itself; as we argued above, that estimate gives information on the rate of backward volume contraction in the unstable direction. Since the noise generally destruct these invariants sets, we expect the extremal index be equal to $1$ or quickly approaching $1$ when the noise increases.
This is confirmed by the numerical experiments reported in Figure \ref{ExtI} where the value of $\theta$ is plotted against the intensity of the noise $\epsilon$ for three maps:  $3x$ mod $1$ map, the Baker map and the Lozi map. In all cases, indeed $\theta \to 1$ for large enough noise. However, with respect to the observables discussed in \cite{faranda2013extreme}, we find some remarkable differences on the intensity of the noise needed to observe changes of the extremal index from the deterministic values: whereas in \cite{faranda2013extreme} we observed significant deviation from the deterministic behavior for very small noise intensities ($\epsilon\geq 10^{-4}$), here we need $\epsilon\geq 10^{-2}$, i.e. only large noise amplitudes perturb the estimates of $D_2$ and $\theta$. This difference can be easily explained: in \cite{faranda2013extreme}, the extremal index was used to explore the local stability at periodic fixed points, where the dynamics is deeply affected even by a small noise. Here, instead, the extremal index tracks a global property that it is stable with respect to small stochastic perturbations. We underline that,  for the Lozi map, we cannot obtain estimates of $\theta$ for noise larger than $0.1$ because the dynamics fall out the basin of attraction.

\section{Discussion and conclusions}
Using extreme value theory, we have introduced a new and efficient way to compute the correlation dimension $D_2$. Moreover, for higher dimensional maps  we  introduced the quantity $q_0$, related to  the expectation of the  inverse of the determinant  of the derivative along the expanding subspace. Therefore, the extremal index $\theta=1-q_0$ is a  {\em measure of the averaged rate of phase space contraction for backward iteration}. Although  this quantity slightly differs from the entropy or from the positive Lyapunov exponent when the expanding subspace has dimension one, it provides an important piece of information on the dynamics of the system.  In fact it can be linked to the global predictability and therefore   {\em considered as a new indicator of the local instability in chaotic systems}. We would like also to emphasize that both $\theta$ and $D_2$ {\em can be computed simultaneously} just by looking at the GEV function and this makes our method quite rapid and economically efficient from a numerical point of view. We have shown that even for short time series of only of $10^4$ iterations, the estimates are robust and consistent with the theoretical expectations. We have also presented a first application of these indicators to climate data proving that the indicators are useful to infer the spatial number of degrees of freedom and the typical time scales of the atmospheric dynamics on the North Atlantic region. Finally we have observed their  sensitivity to the different boundary conditions imposed for the climate simulations analyzed. This implies that the indicators could be useful in characterizing and comparing also different climate datasets as those analyzed in international campaigns.

Our interpretation of $\theta$  together with that on the correlation dimension $D_2$ could be useful also to analyze times series arising from the evolution of chaotic systems. Indeed, these quantities are particularly straightforward to obtain from numerical computations.
Moreover the results obtained can also be used to detect the embedding dimension, namely by replacing the sample  of data with delay vectors of variable lengths; we stress that   computing  the GEV with those delay vectors will allow us to get exactly the embedding dimension. We mean to develop further this approach in a future paper.

Finally, the computation of the DEI  could be helpful to distinguish  purely stochastic sequences for which the extremal index should approach $1,$ see Section 4.4, from dynamical systems with an underlying chaotic behavior even in presence of small stochastic perturbations. Again, these further applications of our approach with EVT will be the objects of forthcoming investigations.

\section{Supplementary Material}
See supplementary material for: i) the algorithm for the estimation of the correlation dimension $D_2$ and the Dynamical Extremal Index (DEI) $\theta$, ii) a commented numerical MATLAB code for such estimation iii) the model equations for the maps used iv) the supplementary figures.

\section*{Acknowledgments}

SV was supported by  the MATH AM-Sud Project Physeco,    by  the Leverhulme Trust  thorough the Network Grant IN-2014-021  and by the project APEX Syst\`emes dynamiques: Probabilit\'es et Approximation Diophantienne
PAD funded by the R\'eegion PACA (France). DF was partially supported by the ERC grant A2C2 (No. 338965). The authors warmly thank the referee whose comments and advices helped them to improve the paper.


%
%
%
%
%

\bibliography{dimbound}%

\clearpage

\begin{table}[]
\centering
\caption{Estimates of correlation dimension $D_2$ and dynamical extremal index (DEI) $\theta$  obtained with $l=100$ trajectories, consisting of $n=10^6$ iterations or $n=10^4$ iterations. Maxima of $\psi(x,y)$ are extracted in block of $s=10^3$ and $s=10^2$ length, for a total of $m=10^3$ or $m=10^2$ blocks. The quantile for the estimate of the DEI is $\tilde{s}=0.99$. For the Arnold Cat's map the convergence to theoretical value is lower and the estimates are provided only for $\tilde{s}=0.99999$ and $n=10^7$.}
\label{table0}
\begin{tabular}{|l|c|c|c|c|c|c|}
\hline
Map                & $D_2$ (classical) & $D_2$ ($n=10^6$) & $D_2$ ($n=10^4$)   & $\theta$ (from Lyapunov) & $\theta$ ($n=10^6$) & $\theta$ ($n=10^4$) \\ \hline
Bernoulli's Shifts & 1                & 1.00$\pm$0.02 & 1.01$\pm$0.14 & 0.667  & 0.668$\pm$0.004 & 0.69$\pm$0.04 \\
Gauss map          & 1                & 1.00$\pm$0.03 & 0.96$\pm$0.16 &	 0.773 & 0.773$\pm$0.005 & 0.78$\pm$0.04 \\
Cantor IFS         & 0.667            & 0.64$\pm$0.01 & 0.59$\pm$0.13 & 0.5                  & 0.502$\pm$0.005 & 0.50$\pm$0.05 \\
Baker map          & 1.41               &  1.46$\pm$0.02  & 1.42$\pm$0.25        & 0.47                 &      0.49$\pm$ 0.02 & 0.50$\pm$0.04 \\
Lozi map           &   1.38               & 1.39$\pm$0.11  & 1.29$\pm$0.25 & 0.37 &   0.37$\pm$ 0.01 & 0.37$\pm$0.05\\
Henon map          &   1.22               & 1.24$\pm$0.03   &1.13$\pm$0.25 & 0.34  &   0.43$\pm$0.01  & 0.43$\pm$0.06  \\
Solenoid $a=1/3$          &   1.6309               & 1.64$\pm$0.04  &1.55$\pm$ 0.17& 0.5                &   0.51$\pm$0.01 & 0.59$\pm$0.03   \\
Solenoid $a=1/4$          &   1.5               & 1.52$\pm$0.03  &1.57$\pm$0.20 & 0.5                 &   0.51$\pm$0.01   & 0.53$\pm$0.03 \\
Arnold Cat's map   & 1.987            & 2.00$\pm$0.06 &  --& 0.51                 & 0.53$\pm$0.06  & --\\
 \hline
\end{tabular}
\end{table}

\begin{table}[]
\centering
\caption{Estimates of correlation dimension $D_2$ and extremal index $\theta$ obtained for daily sea-level pressure maps for four different periods of the  ERA-20CM reanalysis. Values represent average over the 9 ensemble members and uncertainty is expressed as the standard deviation of the ensemble mean.}
\label{table1}
\begin{tabular}{|l|c|c|}
\hline
    Period             & $D_2$  & $\theta$ \\ \hline
1900-2010 & 8.9$\pm$ 0.8 & 0.48$\pm$0.05\\  \hline
1900-1955 & 8.8$\pm$ 0.7 & 0.50$\pm$0.03\\  \hline
1900-1928 & 9.4$\pm$ 0.8 & 0.50$\pm$0.02\\  \hline
1900-1914 & 9.0$\pm$ 1.0 & 0.50$\pm$0.03\\  \hline

\end{tabular}
\end{table}

\end{document}